\documentstyle[12pt]{article}
\begin{document}

\begin{center}

\vspace{5mm}

{\large \bf QCD AND EXPERIMENT ON MULTIPLICITY DISTRIBUTIONS}

\vspace{3mm}

{\large  I.M.\,DREMIN}

\vspace{2mm}

{\normalsize  Lebedev Physical Institute, Moscow 117924, Russia }

\end{center}

\begin{abstract}
The solution of QCD equations for generating functions of {\it parton}
multiplicity distributions reveals new peculiar features of cumulant moments
oscillating as functions of their rank. It happens that experimental
data on {\it hadron} multiplicity distributions in $e^{+}e^{-}, hh, AA$
collisions possess the similar features. However, the "more regular" models
like $\lambda \phi ^3$ behave in a different way. Evolution of the moments at
smaller phase space bins and zeros of the truncated generating functions are
briefly discussed.
\end{abstract}

\section{Introduction and main results}

Since the space-time (space in proceedings and time at oral presentation) is
very limited I present from the beginning both short review of the problem
and the main results obtained, leaving their justification
and brief discussion for the next
section. Those interested in more detailed description should use the list of
references for further reading (in particular, the review paper \cite{1}).

I would like to stress that QCD predicts the distributions of
partons (quarks and gluons) while in experiment one gets the distributions
of final hadrons. Therefore no {\underline \it quantitative} comparison has
been attempted. However, the {\underline \it qualitative} features of both
distributions are so spectacular and remind each other that one is tempted
to confirm once again that QCD is a powerful tool for predicting new features
of hadron distributions as well.

For a long time, the phenomenological approach dominated in description of
multiplicity distributions in multiparticle production \cite{2}. The very first
attempts to apply QCD formalism to the problem failed because in the
simplest double-logarithmic approximation it predicts an extremely wide shape
of the distribution \cite{3} that contradicts to experimental data. Only
recently it became possible \cite{4} to get exact solutions
of QCD equations for the generating functions of multiplicity distributions
which revealed much narrower shapes and such a novel feature
of cumulant moments as their oscillations at higher ranks \cite{5},\cite{6}.
The similar oscillations have been found in experiment for the moments of
hadron distributions \cite{7}. Their pattern differs drastically from those
of the popular phenomenological distributions \cite{1} and of the
"non-singular"
$\lambda \phi ^3$-model \cite{8}.

These findings have several important implications \cite{1}. They show that:

1.{\it the QCD distribution belongs to the class of non-infinitely-divisible
ones.}

Two corollaries of this statement follow immediately: \\
{\it a}. The Poissonian cluster models (e.g., the multiperipheral cluster
model)
are ruled out by QCD. \\
{\it b}. The negative binomial distribution (so popular nowadays in
phenomenological fits) can not be valid asymptotically. \\

2.{\it the new expansion parameter appears in description of multiparticle
processes.}

Since this parameter becomes large when large number of particles are
involved, it asks for the search of some collective effects and of
more convenient basis than the common particle number representation.

QCD is also successful in qualitative description of evolution of multiplicity
distributions with decreasing phase space bins which gives rise to notions
of intermittency and fractality \cite{9},\cite{10},\cite{11}.
However, there are some new problems
with locations of the minimum of cumulants at small bins \cite{12},\cite{13}.

The experimentally defined truncated generating functions possess an intriguing
pattern of zeros in the complex plane of an auxiliary variable \cite{13},
\cite{14},\cite{15}. It recalls
the pattern of Lee-Yang zeros of the grand canonical partition function in
the complex fugacity plane related to phase transition \cite{16},\cite{17}
and asks for some collective effects to be searched for \cite{18},\cite{19}.

\section{Technicalities}
Let us define the multiplicity distribution
\begin{equation}
P_n = \sigma _n /\sum_{n=0}^{\infty }\sigma _n  ,     \label{1}
\end{equation}
where $\sigma _n$ is the cross section of $n$-particle production processes,
and the generating function
\begin{equation}
G(z) = \sum _{n=0}^{\infty }P_{n}(1+z)^n .     \label{2}
\end{equation}
The (normalized) factorial and cumulant moments of the $P_n$ distribution are
\begin{equation}
F_{q} =\frac {\sum_{n} P_{n} n(n-1)...(n-q+1)}{(\sum_{n} P_{n} n)^q} = \frac
{1}
{\langle n\rangle ^q} \frac {d^qG(z)}{dz^q}\vert _{z=0} ,  \label{3}
\end{equation}
\begin{equation}
K_{q} = \frac {1}{\langle n\rangle ^q}\frac {d^q \ln G(z)}{dz^q}\vert _{z=0},
\label{4}
\end{equation}
where $\langle n\rangle = \sum _{n} P_{n} n$ is the average multiplicity. They
describe full and genuine $q$-particle correlations, correspondingly. Let us
point out here that the moments are defined by the derivatives at the origin
and are very sensitive to any nearby singularity of the generating function.

First, let us consider QCD without quarks, i.e. gluodynamics. The generating
function of the gluon multiplicity distribution in the full phase-space
volume satisfies the equation
\begin{equation}
\frac {\partial G(z,Y)}{\partial Y} = \int _{0}^{1}dxK(x)\gamma _{0}^{2}
[G(z,Y+\ln x)G(z,Y+\ln (1-x)) - G(z,Y)] .    \label{8}
\end{equation}
Here $Y=\ln (p\theta /Q_{0}), p$ is the initial momentum, $\theta $ is the
angular width of the gluon jet considered,  $p\theta \equiv Q$
where $Q$ is the jet virtuality, $Q_{0}=$const,
\begin{equation}
\gamma _{0}^{2} = \frac {6\alpha _{S}(Q)}{\pi } ,  \label{9}
\end{equation}
$\alpha _S$ is the running coupling constant, and the kernel of the equation is
\begin{equation}
K(x) = \frac {1}{x} - (1-x)[2-x(1-x)] .    \label{10}
\end{equation}
It is the non-linear integro-differential equation with shifted arguments in
the non-linear part which take into account the conservation laws.
, and with
the initial condition
\begin{equation}
G(z, Y=0) = 1+z ,    \label{11}
\end{equation}
and the normalization
\begin{equation}
G(z=0, Y) = 1 .     \label{12}
\end{equation}
The condition (\ref{12}) normalizes the total probability to 1, and the
condition (\ref{11}) declares that there is a single particle at the very
initial stage.

After Taylor series expansion at large enough $Y$ and differentiation in
eq. (\ref{8}), one gets the differential equation
\begin{equation}
(\ln G(Y))^{\prime \prime }= \gamma _{0}^{2}[G(Y)-1-2h_{1}G^{\prime }(Y)+
h_{2}G^{\prime \prime }(Y)] ,    \label{12a}
\end{equation}
where $h_1 = 11/24; h_2 = (67-6\pi ^2)/36\approx 0.216$, and higher order terms
have been omitted.

Leaving two terms on the right-hand side, one gets the well-known equation
of the double-logarithmic approximation which takes into account the most
singular components. The next term, with $h_1$, corresponds to the modified
leading-logarithm approximation, and the term with $h_2$ deals with
next-to-leading corrections.

The straightforward solution of this equation looks very problematic.
However, it is very simple for the moments of the distribution because
$G(z)$ and $\ln G(z)$ are the generating functions of $F_q$ and $K_q$,
correspondingly, according to (\ref{3}), (\ref{4}). Using this fact, one gets
the solution which looks like
\begin{equation}
H_q = \frac {K_q}{F_q} = \frac {\gamma _{0}^{2}[1-2h_{1}q\gamma +h_{2}(q^2
\gamma ^{2} + q\gamma ^{\prime })]}{q^2 \gamma ^2 + q\gamma ^{\prime }},
\label{13}
\end{equation}
where the anomalous dimension $\gamma $ is related to
$\gamma _0$ by
\begin{equation}
\gamma = \gamma _0 - \frac {1}{2}h_{1}\gamma
_{0}^{2} + \frac {1}{8} (4h_2 - h_{1}^{2})\gamma _{0}^{3} + O(\gamma _{0}^{4})
.   \label{14}
\end{equation}

The formula (\ref{13}) shows how the ratio $H_q$ behaves in different
approximations. In double-log approximation, where $h_1 = h_2 = 0$, it
monotonously decreases as $q^{-2}$ that corresponds to the negative binomial
law with its parameter $k=2$ i.e. to very wide distribution. Let us note that
it owes to the singular part of the kernel and is absent on more regular
theories like $\lambda \phi ^3$. In modified-log
approximation ($h_2 = 0$) it acquires a negative minimum at
\begin{equation}
q_{min}=\frac {1}{h_{1}\gamma _{0}}+\frac {1}{2}+O(\gamma _0) \approx 5
\label{q}
\end{equation}
and approaches the abscissa axis from below asymptotically at large ranks $q$.
In the next approximation given by (\ref{13}) it preserves the minimum location
 but approaches a
positive constant crossing the abscissa axis. In ever higher orders it
reveals the quasi-oscillatory behavior about this axis. This prediction of the
minimum at $q\approx 5$ and subsequent specific oscillations is the main
theoretical outcome.

It is interesting to note that the equation (\ref{8}) can be solved exactly
in the case of fixed coupling constant \cite{6}. All the above qualitative
features are noticeable here as well.

While the above results are valid for gluon distributions in gluon jets
(and pertain to QCD with quarks taken into account \cite{1}),
the similar qualitative features characterize the multiplicity distributions
of hadrons in high energy reactions initiated by various particles.
The numerous demonstration of it can be found in the review paper \cite{1}.

The multiplicity distributions can be measured not only in the total phase
space (as has been discussed above for very large phase-space volumes) but
in any part of it. For the homogeneous distribution of particles within the
volume, the average multiplicity is proportional to the volume and decreases
for  small volumes but the fluctuations increase. The most interesting
problem here is the law governing the growth of fluctuations and its possible
departure from a purely statistical behavior related to the decrease of the
average multiplicity. Such a variation has to be connected with the dynamics of
the
interactions. In particular, it has been proposed to look for the power-law
behavior of the factorial moments for small rapidity intervals $\delta y$
\begin{equation}
F_q \propto (\delta y)^{-\phi (q)} \;\;\;\;(\phi (q)>0)\;\;\;\;
(\delta y \rightarrow 0), \label{15}
\end{equation}
inspired by the idea of intermittency in turbulence. In the case of statistical
fluctuations with purely Poissonian behavior, the intermittency indices $\phi
(q)$ are identically equal to zero.

Experimental data on various processes in a wide energy range support this idea
, and QCD provides a good basis for its explanation as a result
of parton showers.

At moderately small rapidity windows, one can get in the double-log
approximation the power-law behavior with
\begin{equation}
\phi (q) = D(q-1) - \frac {q^2 - 1}{q}\gamma _0 .   \label{16}
\end{equation}
The running property of QCD coupling constant is not important in that region.
This property becomes noticeable at ever smaller windows, where (e.g., at
$q$=2)
$\ln \delta y_0/\delta y >\alpha _{S}^{-1}$, and leads to smaller numerical
values of $\phi (q)$ compared to (\ref{16}). The general trends in this region
decline somewhat from the simple power law (\ref{15}) due to logarithmic
corrections. Qualitatively, these predictions correspond to experimental
findings at relatively small ranks $q$ where the steep increase
in the region of $\delta y>1$
on the log-log plot of the dependence (\ref{15}) is replaced by slower one at
smaller intervals $\delta y$. The transition point between
the two regimes depends on the rank in qualitative agreement with QCD
predictions also. Namely, the transition happens at smaller bins for higher
ranks. These findings can be interpreted as an indication on
fractal structure of particle distributions within the available phase space.
When interpreted in terms of fluctuations, they show that the fluctuations
become stronger in small phase-space regions in a definite power-like
manner and, surely, exceed trivial statistical fluctuations.

Let us turn now to the $q$-behavior of moments at small bins. The phenomenon of
the oscillations of cumulants discussed above reveals itself here as well if
one
goes beyond the double-log approximation of (\ref{16}). In terms of factorial
moments, it means the non-monotonous behavior of the intermittency indices as
functions of $q$. (Compare it to the steady increase with $q$ at $q>1$
given by (\ref{16}).) It gives rise to the negative values of $K_q$ and $H_q$.

The fate of the first minimum can be easily guessed from the formula (\ref{q}).
For large enough virtualities (i.e. small $\gamma _{0}$), the minimum location
moves to higher values of rank $q$ for jets
with larger virtuality $Q$ since the QCD coupling constant is running as
$\ln ^{-1}Q$. Therefore, the predicted shift of the minimum is
\begin{equation}
q_{min}\propto \ln ^{1/2}Q .
\end{equation}
It follows that $q_{min}$ moves to higher ranks at higher energies because
more massive jets become available. Another corollary is that it should
shift to smaller values of $q$ for smaller bins at fixed energy.

While former statement finds some support in experiment, the second one does
not look to be true. On the contrary, the minimum appears
at higher ranks for smaller bins. There is no solution of this problem yet
but it should be ascribed to the higher-order effects.
Actually, one can guess that the higher order terms shown as $O(\gamma _{0})$
in (\ref{q}) become so important at small bins that they overpower the weak
$Q$-dependence of $\gamma _{0}^{-1}$ in the first term of (\ref{q}). It is
important to stress here that at large rapidity intervals the modified
leading-log term with $h_2$ does not influence the value of $q_{min}$, and
only increases the value of $H_q$ by $h_{2}\gamma _{0}^{2}$. Thus, the next-
to-leading corrections should be in charge of the additional shift of
$q_{min}$,
and, therefore, small bins help us look into higher orders of QCD.

There is another fascinating feature of multiplicity distributions -- it
happens
that zeros of the truncated (sum in \ref{2} runs up to $n=N_{max}$) generating
function form a spectacular pattern in
the complex plane of the variable $z$. Namely, they seem to lie close to a
single circle. At enlarged values of $N_{max}$ they move closer to the real
axis pinching it at some positive value of $z$.

No QCD interpretation of the fact exists because it is hard to exploit the
finite cut-off in analytic calculations. The interest to it stems from the
analogy to the locations of zeros of the grand canonical partition function
as described
by Lee and Yang who related them to possible phase transitions in statistical
mechanics. In that case, $z$ variable plays the role of fugacity, and pinching
of the real axis implies existence of two phases in the system considered.

In particle physics, it shows up the location of the singularity of the
generating function i.e. the number of zeros of truncated generating functions
increases and they tend to move to the singularity point when $N_{max}
\rightarrow \infty $. Since it happens to lie close to the origin, it
drastically influences the behavior of moments (see (\ref{3}), (\ref{4})), and,
therefore, determines the shape of the distribution. The study of the
singularities is at the very early stage now, and one can only say that the
singularity is positioned closer to the origin in nucleus-nucleus collisions
and it is farthest in $e^{+}e^{-}$ that appeals to our intuitive guess.

To conclude, I would like to stress that, once again, QCD demonstrates its
power in predicting new features of {\it particle} distributions when
dealing with {\it parton} distributions.

\vspace{1mm}

{\large Acknowledgments}

\vspace{1mm}

I am grateful to J.Tran Thanh Van for inviting me to participate in 2nd
Rencontres du Vietnam and for financial support.

This work is supported by the Russian Fund for Fundamental Research (grant
93-02-3815) and by INTAS grant 930-0079.\\

\end{document}